\documentclass[aps,pra,reprint,groupedaddress,nofootinbib,floatfix]{revtex4-2}

\usepackage{amsmath,amssymb,bm}
\usepackage{graphicx}
\usepackage{booktabs}
\usepackage{xcolor}
\usepackage[colorlinks=true,allcolors=blue!45!black]{hyperref}
\graphicspath{{figures/}}

\newcommand{\E}{\mathbb{E}}
\newcommand{\dd}{\mathrm{d}}
\newcommand{\sinc}{\operatorname{sinc}}

\begin{document}

\title{Finite-event modulation transfer through forward-scattering media}

\author{Chong Wang}
\author{Liangfeng Hu}
\author{Yuchen He}
\email{yuchenhe@xjtu.edu.cn}
\author{Huaibin Zheng}
\author{Hui Chen}
\author{Jianbin Liu}
\affiliation{Electronic Materials Research Laboratory, Key Laboratory of the Ministry of Education and International Center for Dielectric Research, School of Electronic Science and Engineering, Xi'an Jiaotong University, Xi'an 710049, People's Republic of China}

\author{Yu Zhou}
\affiliation{MOE Key Laboratory for Nonequilibrium Synthesis and Modulation of Condensed Matter, Department of Applied Physics, Xi'an Jiaotong University, Xi'an 710049, People's Republic of China}

\author{Zhuo Xu}
\affiliation{Electronic Materials Research Laboratory, Key Laboratory of the Ministry of Education and International Center for Dielectric Research, School of Electronic Science and Engineering, Xi'an Jiaotong University, Xi'an 710049, People's Republic of China}

\date{Draft compiled \today}

\begin{abstract}
How spatial structure is transferred through a forward-scattering medium depends jointly on pattern scale and propagation distance. Starting from photon exit displacements, we derive modulation transfer while retaining the number, depths, and projected angles of individual deflections, with event occurrences described by a homogeneous Poisson process along the propagation path. The second angular moment yields a leading low-frequency loss proportional to $k^2z^3$, where $k$ is angular spatial frequency and $z$ is thickness. At finite frequency, the model predicts two coupled departures: loss bends below its low-frequency quadratic extrapolation at fixed thickness, and transfer curves separate by thickness at fixed $k^2z^3$. The leading corrections to both effects arise from the same fourth angular moment. For symmetric, nondegenerate projected-angle distributions, the full response further establishes increasing loss with thickness in the same medium at fixed joint variable. Binary-pattern measurements through TiO$_2$ suspensions span five nominal loadings, seven thicknesses, and 20 spatial frequencies, supporting the approximate low-frequency scaling and showing both departures. A Gaussian effective-angle response calibrated only on lower-frequency data predicts the held-out higher-frequency band without refitting, with a logarithmic root-mean-square error of 0.0181 compared with 0.1021 for quadratic extrapolation. The results show that finite-event statistics provide a common physical origin for spectral bending and thickness-dependent departure from low-frequency scaling, linking coarse-pattern transport to finer spatial structure.
\end{abstract}

\maketitle

\section{Introduction}

Structured intensity patterns---including sinusoidal fringes, binary masks, and other deterministic modulations---are widely used to encode optical measurements and recover spatial information \cite{Neil1997,Shapiro2008,Wang2026Graph}. They support optical sectioning and spatial-frequency-domain characterization \cite{Cuccia2005,Cuccia2009,Kanick2014,Hu2017,Gioux2019}, as well as computational imaging through scattering media \cite{Zhao2022,Wang2025Implicit}. When these patterns propagate through a scattering medium, their spatial modulation is progressively attenuated. Fine patterns generally lose contrast more readily than coarse patterns, and propagation thickness further changes their attenuation. Understanding how pattern transfer depends on spatial frequency and medium thickness therefore provides a basis for interpreting these measurements and for predicting finer-pattern transfer from coarse-pattern data.

Small-angle transport theory relates scattering-induced beam spreading to image modulation transfer \cite{Wells1969,Yura1971,Benke1990,McLean1991,Ishimaru1978}. Subsequent studies have examined the spatial-frequency dependence of the transfer function, pattern transmission through layered media, and the contributions of different scattering orders \cite{Zardecki1984,Kuga1985,Kuga1986,Gray2012}. Numerical transport calculations and analyses of small-angle approximations have also examined imaging geometry and the range of validity of these descriptions \cite{Bruscaglioni1993,Jaffe1995,Sassaroli1998,Swanson2001}. In the angular-diffusion description of a statistically homogeneous medium, the variance of the transverse exit displacement grows as the cube of the propagation thickness, yielding the normalized modulation transfer \cite{Benke1990,Dolin2013}
\begin{equation}
 H(k,z)\simeq \exp(-\alpha k^2z^3),
 \label{eq:diffusion-reference}
\end{equation}
where $k$ is angular spatial frequency and $z$ is medium thickness. This relation gives a simple low-frequency scaling: frequency--thickness pairs with the same joint variable $X=k^2z^3$ have the same modulation loss. The leading low-frequency loss is determined by the exit-displacement variance, whereas transfer at finite frequency also depends on higher-order statistics of the displacement distribution. We examine how these statistics jointly affect the spectral shape and pattern transfer at different thicknesses, and whether low-frequency measurements can predict modulation loss at higher spatial frequencies.

\begin{figure*}[!tbp]
 \includegraphics[pagebox=cropbox,width=\textwidth]{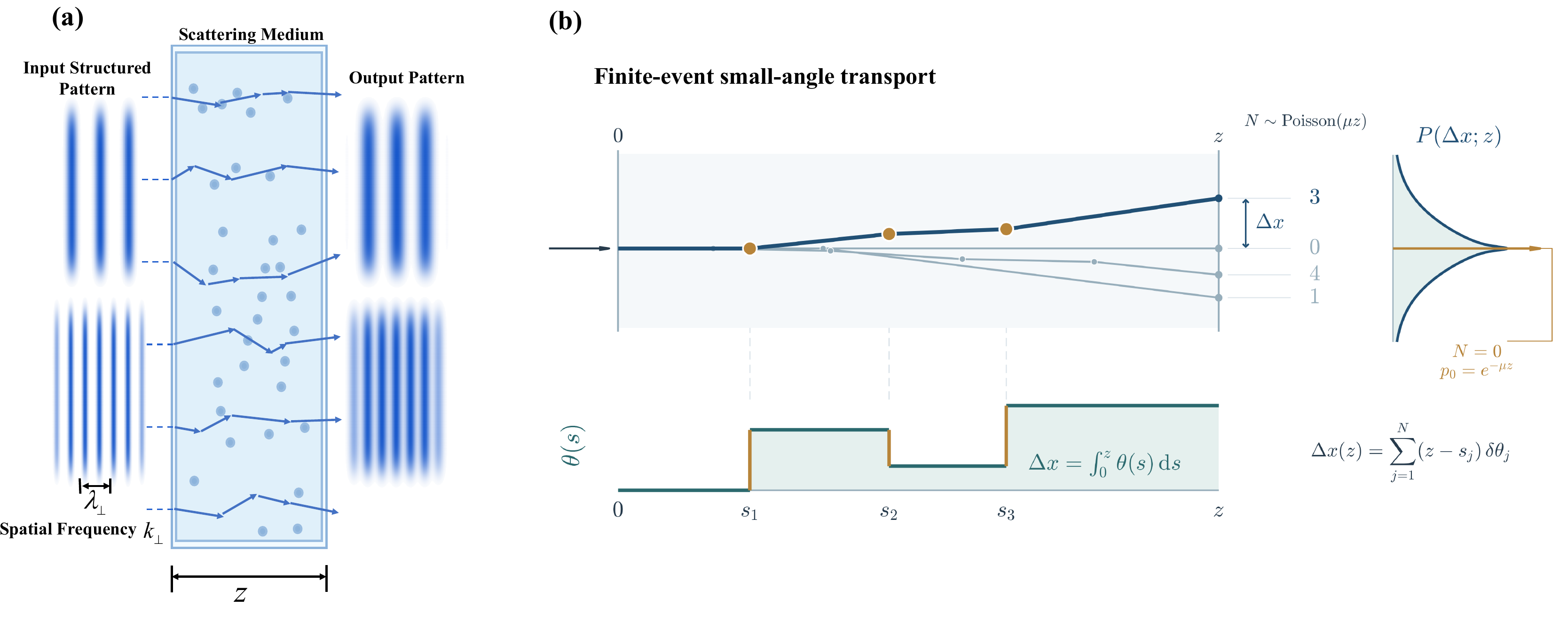}
 \caption{\label{fig:physical-picture}Scale-dependent transfer of structured intensity patterns and schematic finite-event small-angle transport. (a) Fine stripes lose more contrast than coarse stripes after the same medium. (b) Illustrative paths undergo $N=0,1,3,4$ deflections; the model takes $N\sim\mathrm{Poisson}(\mu z)$. Event $j$ at depth $s_j$ adds a projected angle $\delta\theta_j$. The accumulated angle $\theta(s)$ gives $\Delta x=\sum_{j=1}^{N}(z-s_j)\delta\theta_j$. The path ensemble defines $P(\Delta x;z)$; the gold marker at $\Delta x=0$ has zero-event probability $p_0=e^{-\mu z}$.}
\end{figure*}

Here, we derive pattern transfer through a forward-scattering medium from the transverse displacement of photon paths, retaining the number, positions, and angles of individual scattering events. The model describes event occurrences by a homogeneous Poisson process along the propagation path. The second angular moment gives the low-frequency $k^2z^3$ scaling, while the fourth moment determines the leading corrections responsible for both spectral bending and thickness layering at fixed $X$. For symmetric, nondegenerate projected-angle distributions, the model also establishes that modulation loss increases monotonically with thickness in the same medium at fixed $X$. Stripe-transmission measurements through TiO$_2$ suspensions at five nominal loadings and seven thicknesses test the low-frequency scaling, spectral bending below the quadratic extrapolation, and thickness layering. Parameters determined from low-frequency data then predict the held-out higher-frequency data without refitting. These results connect spectral bending and thickness layering to a common scattering-event mechanism and extend coarse-pattern scaling to predictions of finer-pattern transfer. Figure~\ref{fig:physical-picture} illustrates scale-dependent pattern transfer and the scattering-path geometry.

\section{Finite-event small-angle transfer model}

To calculate modulation transfer beyond the low-frequency approximation, we retain the number, depths, and projected angles of the deflection events. We assume paraxial propagation with parallel incidence, statistical homogeneity along and across the slab, a depth-independent event rate and angular law, and complete collection of transmitted light. Below, $\theta_j$ is the projected increment $\delta\theta_j$ in Fig.~\ref{fig:physical-picture}b.

\subsection{From one deflection to an exit displacement}

A deflection at depth $s$ shifts the exit position of a photon entering at $x_0$ by
\begin{equation}
 \Delta x=(z-s)\theta .
 \label{eq:single-lever}
\end{equation}
Earlier deflections act over a longer remaining distance. For a path with $N(z)$ events at depths $s_j$ and projected angles $\theta_j$, the exit position is $x_{\rm out}=x_0+X_z$, where
\begin{equation}
 X_z=\sum_{j=1}^{N(z)}(z-s_j)\theta_j .
 \label{eq:path}
\end{equation}
This paraxial relation defines the random displacement $X_z$ from the event count, depths, and angles.

\subsection{From photon statistics to arbitrary-pattern transfer}

Independent photon paths generate the normalized displacement measure $P_z(\dd\xi)=\Pr(X_z\in\dd\xi)$. In the Poisson model, paths without deflections retain zero displacement and occur with probability $e^{-\mu z}$. The displacement distribution therefore includes this zero-event contribution concentrated at zero displacement. With transverse statistical homogeneity and complete collection, displaced incident intensities add at the exit plane. Any input pattern therefore obeys
\begin{equation}
 I_{\rm out}(x,z)=T_0(z)\int_{\mathbb R}
 I_{\rm in}(x-\xi)\,P_z(\dd\xi),
 \label{eq:convolution}
\end{equation}
Here $T_0(z)$ collects frequency-independent losses. Point-spread-function descriptions similarly relate image formation in turbid media to the distribution of transmitted light \cite{Gan1999,Rogers2019}. Fourier transformation converts the convolution into multiplication \cite{Goodman2017}; dc normalization removes $T_0$ and leaves the transfer of each frequency:
\begin{equation}
 H(k,z)=\int_{\mathbb R}e^{\mathrm{i}k\xi}P_z(\dd\xi)
       =\E[e^{\mathrm{i}kX_z}].
 \label{eq:characteristic}
\end{equation}
This characteristic function includes both the continuous displacement distribution and the contribution of zero-event paths. Symmetry makes $H$ real; the symmetric Poisson model below also makes it positive.

\subsection{Averaging angles, event depths, and event count}

Compound Poisson processes have been used to describe angular multiple scattering in random media \cite{LeBihan2009}. Here, effective deflections in disjoint short segments are independent. In a segment of length $\dd s$, one event occurs with probability $\mu\,\dd s+o(\dd s)$ and multiple events with probability $o(\dd s)$, giving a homogeneous Poisson count. Conditional on $N(z)=n$, the event depths are independent and uniform on $[0,z]$ when arbitrarily labeled; depths arranged in propagation order are their order statistics \cite{Kingman1992}. The projected angles $\theta_j$ are independent, identically distributed, and independent of the depths. Let $\phi(t)=\E[e^{\mathrm{i}t\theta}]$. Averaging the phase factor of an event at depth $s$ over its angle gives $\phi[k(z-s)]$; averaging over its depth then gives
\begin{equation}
 q(k,z)=\frac{1}{z}\int_0^z\phi[k(z-s)]\,\dd s.
 \label{eq:one-event-average}
\end{equation}
Independence gives $H_n=q^n$ at fixed $n$.  Finally, averaging over $N\sim{\rm Poisson}(\mu z)$ yields
\begin{equation}
 H=e^{-\mu z}\sum_{n=0}^{\infty}\frac{(\mu zq)^n}{n!}
   =e^{\mu z(q-1)}.
 \label{eq:poisson-average}
\end{equation}
Substitution and $u=z-s$ give
\begin{equation}
 H(k,z)=\exp\!\left[-\mu\int_0^z
 \{1-\phi(ku)\}\,\dd u\right].
 \label{eq:master-complex}
\end{equation}
For a symmetric projected-angle law, $\phi(t)=\E[\cos(t\theta)]$.  The real exponent above is nonpositive, hence $H>0$.  Define $L=-\ln H$ and integrate the fixed-angle cosine over the remaining distance $u$:
\begin{equation}
 \boxed{\begin{aligned}
 L(k,z)&=\mu\int_0^z \E[1-\cos(ku\theta)]\,\dd u\\
 &=\mu z\,\E[1-\sinc(kz\theta)] .
 \end{aligned}}
 \label{eq:master}
\end{equation}
Here $\sinc t=\sin t/t$ and $\sinc0=1$. Averaging an event's remaining propagation distance produces the sinc factor, with $\theta$ projected along the pattern wave vector. Equations~(\ref{eq:master-complex}) and (\ref{eq:master}) have the form of small-angle modulation-transfer integrals \cite{Benke1990,Dolin2013,Gray2012}.

\subsection{Low-frequency limit and finite-frequency correction}

Expanding $1-\sinc t$ in Eq.~(\ref{eq:master}) with $t=kz\theta$ gives $1-\sinc t=t^2/6-t^4/120+t^6/5040-\cdots$. Averaging over the random projected single-event angle $\theta$ then introduces $\E(\theta^2)$, $\E(\theta^4)$, and $\E(\theta^6)$. These are the second, fourth, and sixth moments of the angular distribution, denoted by $m_n=\E(\theta^n)$. With finite sixth moment $m_6$, the loss at fixed $z$ becomes
\begin{equation}
 L(k,z)=\frac{\mu m_2}{6}k^2z^3
 -\frac{\mu m_4}{120}k^4z^5
 +\frac{\mu m_6}{5040}k^6z^7+o(k^6),
 \label{eq:series}
\end{equation}
With only $m_4<\infty$, the first two terms still hold with an $o(k^4)$ remainder. The second moment gives the leading low-frequency term in Eq.~(\ref{eq:diffusion-reference}), with $\alpha=\mu m_2/6$. Its $z^3$ factor follows directly from the exit-displacement variance:
\begin{equation}
 \mathrm{Var}(X_z)=\mu m_2\int_0^z(z-s)^2\,\dd s
 =\frac{\mu m_2z^3}{3}.
 \label{eq:displacement-variance}
\end{equation}
Thus $L\simeq k^2\mathrm{Var}(X_z)/2$ at low frequency: events can occur anywhere along the slab, and earlier deflections have a longer lever arm. Continuous angular diffusion is the many-weak-event limit and connects the event description to Fokker--Planck transport \cite{Ishimaru1978,Kim2004}. Holding $\mu m_2=2D_\theta$ fixed as events become more frequent and weaker, while $\mu m_{2n}\to0$ for $n\geq2$, gives at fixed $k,z$
\begin{equation}
 H(k,z)\longrightarrow
 \exp[-D_\theta k^2z^3/3].
 \label{eq:diffusion-limit}
\end{equation}

The fourth moment links two finite-frequency effects. Its negative term bends $L$ below the low-frequency quadratic extrapolation at fixed $z$. At fixed $X=k^2z^3$, it also separates thicknesses: $k=\sqrt{X/z^3}$, so the phase in Eq.~(\ref{eq:master}) becomes $kz\theta=\theta\sqrt{X/z}$. Writing $G_z(X)=L(\sqrt{X/z^3},z)$, Eq.~(\ref{eq:series}) gives, for $X/z\to0$,
\begin{equation}
 G_z(X)=\alpha X-\frac{\beta X^2}{z}
 +\frac{\gamma X^3}{z^2}+o(X^3/z^2).
 \label{eq:joint-expansion}
\end{equation}
Here $\beta=\mu m_4/120$ and $\gamma=\mu m_6/5040$. The leading loss depends only on $X$; the $m_4$ term retains a $1/z$ dependence. Thus the first spectral bend and the first fixed-$X$ thickness separation arise from the same fourth angular moment. The expansion describes the low-frequency region, while Eq.~(\ref{eq:master}) determines the full response.

To compare thicknesses in the same medium, keep $\mu$ and the projected-angle law unchanged and independent of depth. Differentiating Eq.~(\ref{eq:master}) at fixed $X$ gives
\begin{align}
 \left.\frac{\partial G_z}{\partial z}\right|_X
 &=\mu\E[h(w)],\\
 h(w)&=1+\frac{1}{2}\cos w-\frac{3}{2}\sinc w,
 \qquad w=\theta\sqrt{X/z}.
 \label{eq:ordering-derivative}
\end{align}
The even function $h(w)$ is positive for $w\neq0$ (see the Appendix). Consequently, if $\Pr(\theta\neq0)>0$,
\begin{equation}
 \boxed{z_2>z_1\ \Longrightarrow\ G_{z_2}(X)>G_{z_1}(X),\qquad X>0.}
 \label{eq:ordering}
\end{equation}
If $m_2$ is finite, $1-\cos t\leq t^2/2$ further gives
\begin{equation}
 0\leq G_{z_1}(X)<G_{z_2}(X)\leq \alpha X.
 \label{eq:bound}
\end{equation}
At fixed $X$, increasing thickness lowers the pattern frequency as $k=\sqrt{X/z^3}$. The mean event count $\mu z$ increases, while the phase contribution of each event has the decreasing scale $kz\theta=\theta\sqrt{X/z}$. The total phase variance $\mathrm{Var}(kX_z)=\mu m_2X/3$ remains constant: thicker samples combine more events, each with a smaller phase contribution. Equal variance can therefore coexist with different phase distributions and different transfer. For finite $m_2$, the accumulation of these smaller contributions approaches a Gaussian phase distribution as $z$ grows at fixed $X$, and the loss approaches the diffusion line $\alpha X$ from below.

Equation~(\ref{eq:master}) also predicts a high-frequency asymptote. If the projected-angle law assigns zero probability to exactly zero deflection, then $\E[\sinc(kz\theta)]\to0$ as $k\to\infty$ at fixed $z>0$, and
\begin{equation}
 L(k,z)\longrightarrow\mu z,\qquad
 H(k,z)\longrightarrow e^{-\mu z}.
 \label{eq:high-frequency-limit}
\end{equation}
This plateau is the contribution of zero-event paths.

\section{Experimental methods}

\subsection{Pattern generation, samples, and imaging}

\begin{figure}[!t]
 \centering
 \includegraphics[width=\columnwidth]{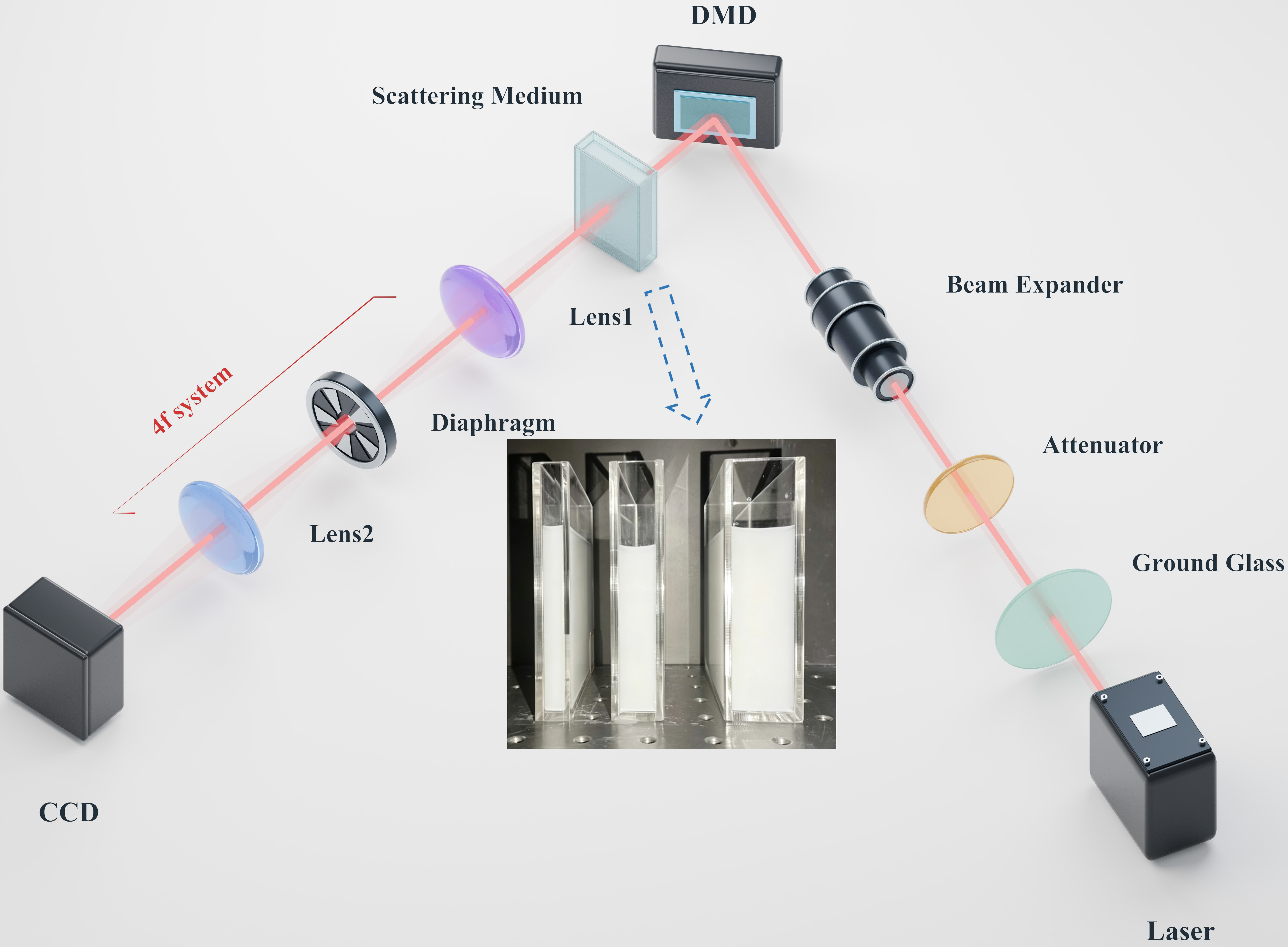}
 \caption{\label{fig:setup}Experimental arrangement for stripe-transfer measurements. Binary stripes generated by the DMD pass through the TiO$_2$ suspension and are imaged onto the CCD. The inset shows acrylic sample cells.}
\end{figure}

A 650-nm laser illuminated a digital micromirror device (UPOLabs HDSLM756D65) through a ground-glass diffuser, variable attenuator, and beam expander (Fig.~\ref{fig:setup}). The device displayed one-dimensional binary square waves with gray levels 0 and 255 and a 50\% duty cycle. Across 1600 active micromirrors, the 20 patterns contained 1--20 periods. After traversing the scattering sample, the projected pattern was relayed to a CCD camera by a unit-magnification $4f$ system with two lenses of focal lengths $f_1=f_2=100$ mm and a diaphragm between them.

The samples were aqueous suspensions of commercial TiO$_2$ powder with a nominal particle diameter of 300 nm. Five nominal powder masses, 0.100, 0.125, 0.150, 0.175, and 0.200 g, were each prepared in 800 mL of water. Seven separate acrylic cells provided nominal propagation thicknesses of 9, 11, 13, 16, 21, 25.5, and 32.5 mm. The suspension was remixed before each repeated acquisition to reduce settling variations. Thickness comparisons therefore use distinct cells prepared at the same nominal loading.

\subsection{Transfer-function extraction}

For each stripe image, we fitted the dc level, a slowly varying linear background, and sine and cosine components at the calibrated fundamental frequency. We normalized the modulation amplitude $A$ by the corresponding dc level and used an exposure-matched air image as the system reference:
\begin{equation}
 H(k,z)=\frac{(A/I_{\mathrm{dc}})_{\mathrm{sample}}}
 {(A/I_{\mathrm{dc}})_{\mathrm{air}}}.
 \label{eq:experimental-H}
\end{equation}
Air images calibrated the physical frequency $f$ at the sample plane; the angular spatial frequency was $k=2\pi f$.

Fixed-$X$ thickness comparisons use the absolute, air-referenced $H$ from separate cells at the same nominal loading. The held-out frequency test instead uses each spectrum's index-4-referenced difference [Eq.~(\ref{eq:D-definition})].

Each loading--thickness--frequency condition was acquired three times and processed independently.  The full acquisition contained 20 frequency indices spanning approximately 0.105--1.473 cycles/mm.  Indices 1--3 were excluded from quantitative spectral analysis because fewer than three periods fell in the analysis region.  The remaining indices 4--20 corresponded to $k\simeq1.844$--9.255 rad/mm.  Index 4 was used as the reference $k_0$ for differential spectra.  Lower-frequency training used indices 5--10; indices 11--20 were withheld for the frequency-extrapolation test.

The 0.150-g data were acquired in a separate batch with multiple air references; the thickest 0.200-g conditions required altered exposure settings. Detailed experimental parameters and transfer-extraction methods are provided in Sec.~S1 and Table~S1 of the Supplemental Material.

\begin{figure}[!t]
 \centering
 \includegraphics[width=\columnwidth]{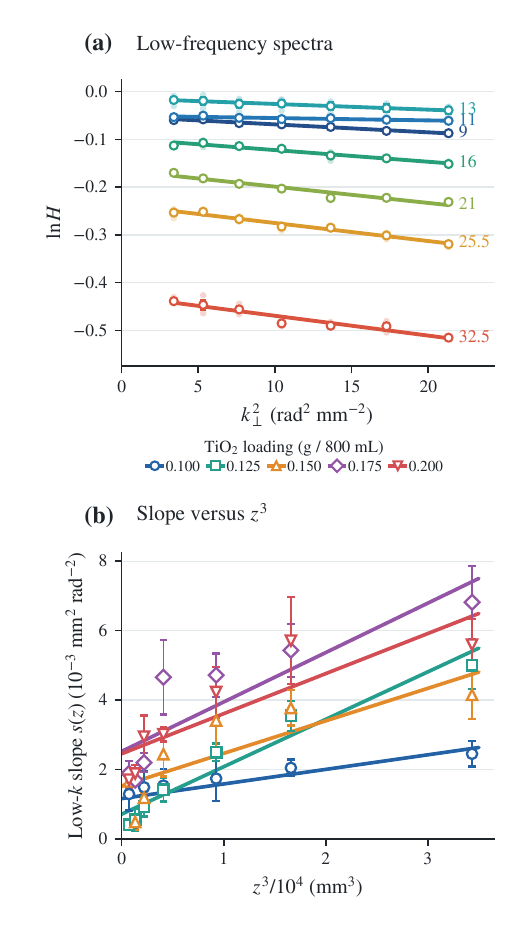}
 \caption{\label{fig:low-frequency}Experimental test of the low-frequency $k^2z^3$ scaling. (a) Low-frequency spectra (indices 4--10) for the 0.150-g/800-mL suspension at seven thicknesses. Pale points show individual measurements; open circles and error bars show the mean and standard error of three repeats. Solid lines are linear fits; numbers at the right give thickness in millimeters. (b) Low-frequency slopes versus $z^3$ for all five loadings. Error bars are 95\% confidence intervals; solid lines are fits of $s(z)=\eta+\beta_{\mathrm{eff}}z^3$.}
\end{figure}

\begin{figure}[!tb]
 \centering
 \includegraphics[width=\columnwidth]{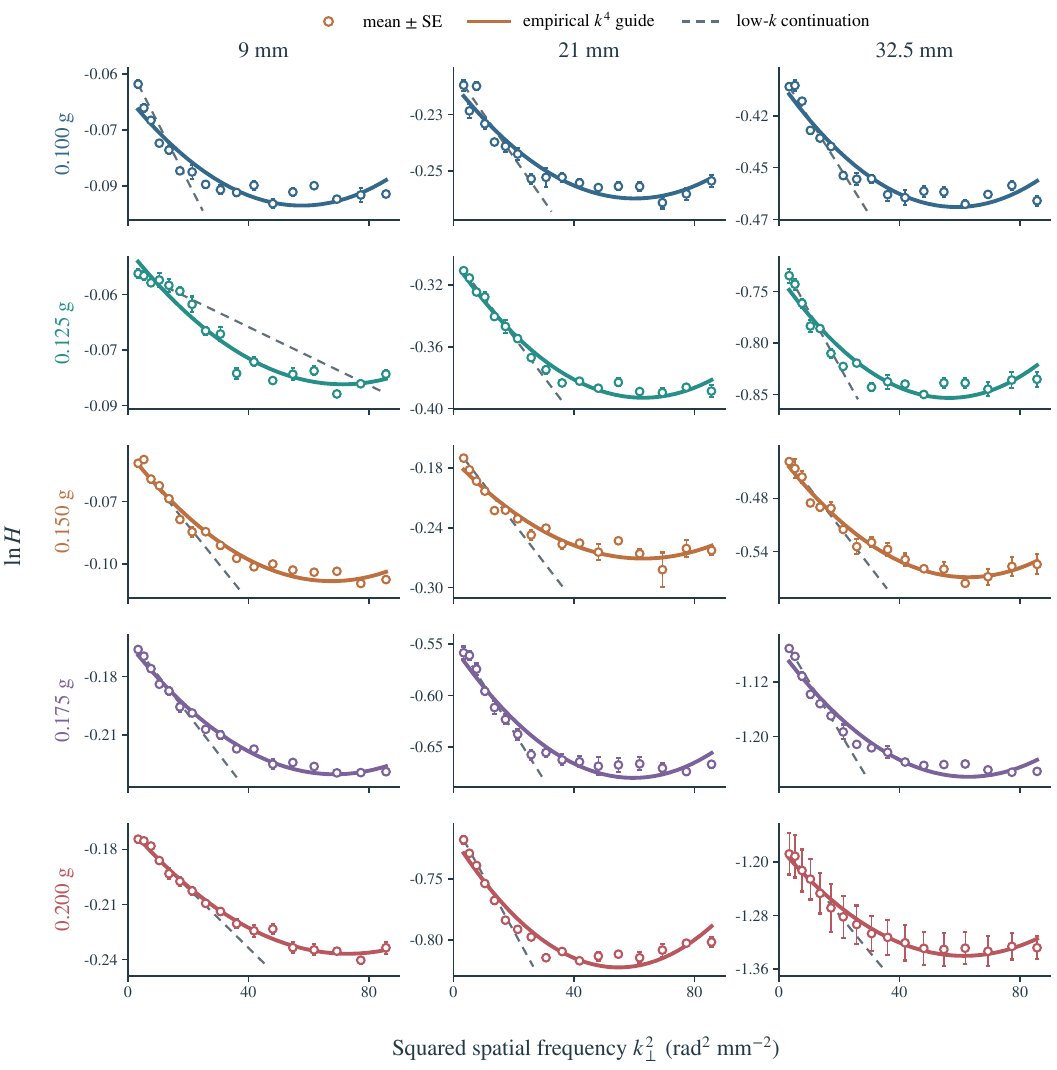}
 \caption{\label{fig:finite-frequency}Full-spectrum bending relative to low-frequency quadratic extrapolation. Rows correspond to five loadings; columns correspond to thicknesses of 9, 21, and 32.5~mm. Open circles and error bars show the mean and standard error of three repeats. Colored solid curves are empirical fits of $a+bk_\perp^2+ck_\perp^4$; gray dashed lines are low-frequency quadratic extrapolations. Vertical ranges differ between panels. All 35 spectra are shown in Supplemental Material, Fig.~S11.}
\end{figure}

\begin{figure*}[!t]
 \centering
 \includegraphics[width=0.82\textwidth]{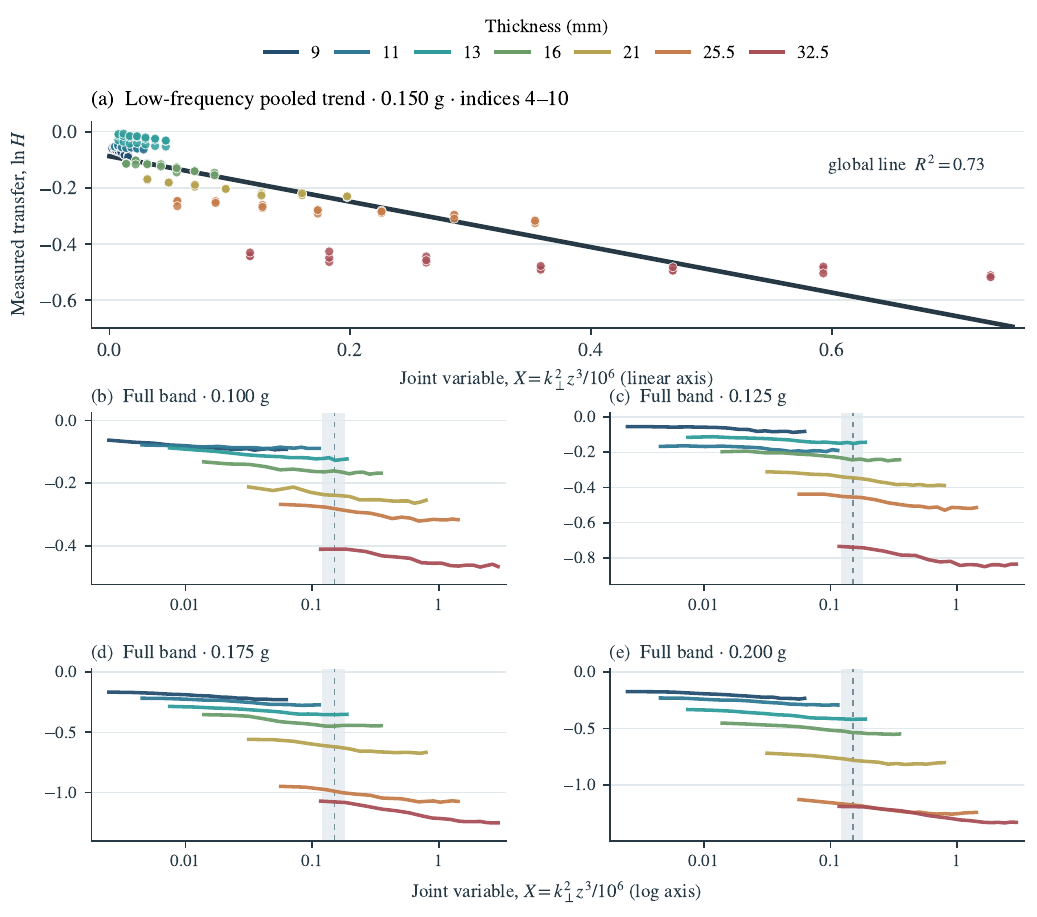}
 \caption{\label{fig:joint-variable}Low-frequency trend and thickness layering at joint variable $X=k_\perp^2z^3$. (a) Low-frequency measurements for the 0.150-g/800-mL suspension at seven thicknesses; the black line is a global linear fit. (b)--(e) Full-band mean curves for the other four loadings, with thickness encoded by color. Shading marks the comparison interval shared by the five thicknesses from 13 to 32.5~mm.}
\end{figure*}

\begin{figure*}[!t]
 \centering
 \includegraphics[width=0.84\textwidth]{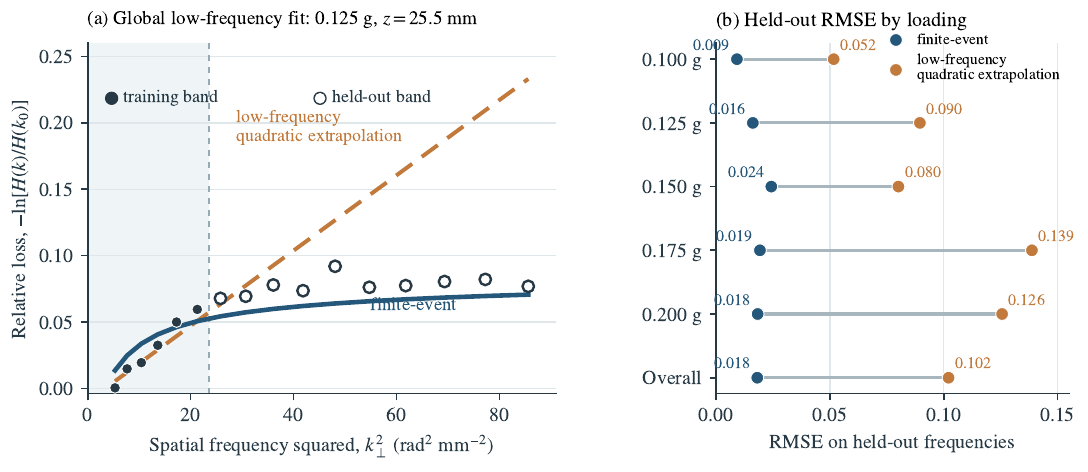}
 \caption{\label{fig:heldout}Held-out higher-frequency predictions from global low-frequency fits. (a) Differential loss for the 0.125-g/800-mL suspension at $z=25.5$~mm. Filled points are training data (indices 5--10); open points are held-out data (indices 11--20). The blue solid curve is the finite-event response and the orange dashed curve is the low-frequency quadratic extrapolation, both using globally fitted parameters. (b) Held-out logarithmic RMSE for each loading and overall.}
\end{figure*}

\subsection{Cross-frequency prediction model and differential loss}

To predict spectra quantitatively, we use a Gaussian projected single-event angle with variance $v$ and characteristic function $\phi(q)=\exp(-vq^2/2)$. Event counts and depths retain the finite-event structure; the spectral bend and fixed-$X$ ordering follow for symmetric angular laws under their stated conditions. Setting $\chi=kz\sqrt{v/2}$ and $t=u/z$ in Eq.~(\ref{eq:master-complex}) gives
\begin{align}
 L(k,z)&=\mu zF(kz;v),\\
 F(kz;v)&=\int_0^1[1-e^{-v(kz)^2t^2/2}]\,\dd t\nonumber\\
 &=1-\frac{\sqrt{\pi}}{2\chi}\operatorname{erf}(\chi).
 \label{eq:gaussian-kernel}
\end{align}
The kernel gives a smooth, bounded crossover, with low-frequency expansion $F=(kz)^2v/6-(kz)^4v^2/40+\cdots$. The fitted $\mu$ and $v$ describe effective transfer.

The prediction test uses differential loss relative to the reference frequency $k_0$:
\begin{equation}
 D_z(k,k_0)=-\ln\!\left[\frac{H(k,z)}{H(k_0,z)}\right]
 =L(k,z)-L(k_0,z).
 \label{eq:D-definition}
\end{equation}
This ratio removes factors common to both frequencies. Parameters are fitted at indices 5--10 and then held fixed for predictions at indices 11--20.

\section{Results}

\subsection{Experimental test of the lower-frequency $k^2z^3$ scaling}

At lower spatial frequencies, each fixed-thickness spectrum falls approximately linearly with $k^2$. For indices 4--10, we write
\begin{equation}
 \ln H(k,z)=a(z)-s(z)k^2,
 \label{eq:low-k-fit}
\end{equation}
The positive slope $s(z)$ measures scale-selective loss. For the representative 0.150-g/800-mL suspension, this relation describes all seven thicknesses [Fig.~\ref{fig:low-frequency}a].

The slopes grow with thickness as $s(z)=\eta+\beta_{\mathrm{eff}}z^3$ [Fig.~\ref{fig:low-frequency}b]. All five loadings give positive $\beta_{\mathrm{eff}}$ ($R^2=0.671$--0.941; $p=0.00030$--0.0242). Together, the nearly linear $k^2$ spectra and their $z^3$-dependent slopes show an approximate $k^2z^3$ scaling over the measured lower-frequency range. The complete low-frequency fits for all five loadings are provided in Figs.~S1--S10 of the Supplemental Material, with slope parameters in Table~S2.

\subsection{Full-spectrum bending beyond the low-frequency quadratic extrapolation}

The finite-event response grows more slowly than its initial quadratic slope as frequency increases. With the Gaussian angular kernel of Eq.~(\ref{eq:gaussian-kernel}), single-event loss begins quadratically in $kz\sqrt{v}$ and gradually levels relative to that continuation. The full spectra test this predicted bend; the held-out band provides a separate prediction test in Fig.~\ref{fig:heldout}.

Fine-pattern loss is lower than the low-frequency quadratic extrapolation predicts. In most loading--thickness panels, measured $\ln H$ rises above the downward continuation as frequency increases [Fig.~\ref{fig:finite-frequency}]. The bend recurs across loadings and thicknesses.

\subsection{Thickness layering in the joint-variable representation}

The same finite-frequency correction that bends a spectrum at fixed thickness also separates thickness curves at fixed $X=k_\perp^2z^3$. In $G_z(X)=\alpha X-\beta X^2/z+\cdots$, the leading term is shared by equal-$X$ data, while the negative correction is larger in magnitude for thinner samples. Comparing thicknesses at fixed $X$ therefore tests a complementary consequence of the spectral bend. The full response predicts greater loss for thicker samples [Eq.~(\ref{eq:ordering})].

The lower-frequency 0.150-g data follow a dominant linear trend in $k_\perp^2z^3$ ($R^2=0.728$; Fig.~\ref{fig:joint-variable}a). Across the full band, however, the other four loadings retain distinct thickness curves in the shaded common interval [Figs.~\ref{fig:joint-variable}b--e], with thicker samples generally giving lower $\ln H$. We interpolated the three-acquisition means at five levels spanning $X/10^6=0.12$--0.18. Absolute loss increased across 13, 16, 21, 25.5, and 32.5~mm in all 25 loading--$X$ comparisons. Thus $X$ captures the dominant trend, while the residual thickness direction matches Eq.~(\ref{eq:ordering}) and complements the fixed-thickness spectral bend.

\subsection{Prediction at held-out frequencies}

Can a response calibrated on coarser patterns predict the bend at finer scales? We referenced each spectrum to index 4 and fitted indices 5--10 across all loadings and thicknesses. The finite-event fit used five loading-dependent amplitudes $\mu_m$ and one shared variance $v$; the low-frequency quadratic extrapolation used five loading-dependent coefficients. Both minimized the same unweighted residual sum of squares over 210 training points. With parameters locked, we predicted all 350 points at indices 11--20 [Fig.~\ref{fig:heldout}a]. These points form 35 reference-normalized spectra; within each spectrum, training and test differences share the measured index-4 reference.

The finite-event response predicts the held-out band with an aggregate logarithmic root-mean-square error of 0.0181, compared with 0.1021 for low-frequency quadratic extrapolation---an 82.3\% reduction [Fig.~\ref{fig:heldout}b]. It has lower error at every nominal loading (0.0091--0.0242 versus 0.0516--0.1388) and in 31 of 35 spectra. Excluding the 32.5-mm conditions gives aggregate errors of 0.0149 and 0.0462, respectively.

Spectral bending alone also admits a simpler bounded description. Under the same protocol, a six-parameter bounded $k$-only response reaches RMSE 0.0263, against 0.0181 for the Gaussian finite-event response; the latter has lower error in 21 of 35 spectra. The thickness direction distinguishes the responses more clearly. At fixed $X$, the fitted $k$-only form predominantly predicts decreasing loss across the shared interval. Measured absolute loss instead rises in all 25 interpolated comparisons, in the direction predicted by finite-event transfer. Together, the frequency and thickness trends favor a propagation-dependent $kz$ response over saturation governed by $k$ alone. Laplace finite-event and empirical bounded-$kz$ forms have held-out RMSEs of 0.0151 and 0.0184, respectively. Model definitions are given in Sec.~S3.1 of the Supplemental Material; held-out predictions are compared in Fig.~S12 and Table~S3.

\section{Conclusions and discussion}

The finite-event formulation describes pattern transfer through photon exit displacements while retaining the count, depth, and angle of individual deflections. Its leading low-frequency loss scales as $k^2z^3$, while the full response links spectral bending and thickness separation. The TiO$_2$ measurements show all three features: approximately quadratic lower-frequency loss with slopes that grow with $z^3$, higher-frequency loss below quadratic extrapolation, and distinct thickness curves at fixed $k^2z^3$. Parameters fitted at indices 5--10 predict the held-out indices 11--20 with a logarithmic loss RMSE of 0.0181, compared with 0.1021 for low-frequency quadratic extrapolation.

The physical connection lies in how different pattern scales probe the exit-displacement distribution. Coarse patterns mainly probe its variance, giving the leading quadratic loss. Finer patterns probe more of the distribution shaped by the deflection events and their remaining propagation distances. Near low frequency, the fourth angular moment produces both the first spectral bend at fixed thickness and the first thickness separation at fixed $X$. The full response carries this connection beyond the low-frequency expansion.

Frequency and thickness provide complementary views of the same transport process. A frequency scan at fixed thickness probes one exit-displacement distribution at different pattern scales. A thickness scan at fixed $X=k^2z^3$ compares phase distributions with the same variance but different event counts and single-event phase scales. The observed spectral bend and thickness ordering thus test two linked consequences of finite-event propagation. Together, they connect coarse-pattern scaling to finer-pattern transfer through the geometry of the photon paths.

\appendix
\section{Positivity underlying the fixed-$X$ ordering}

The sign in Eq.~(\ref{eq:ordering-derivative}) follows from the positivity of $h(w)$. Because $h$ is even, consider $w>0$ and define
\begin{equation}
 g(w)=2w h(w)=w(2+\cos w)-3\sin w.
\end{equation}
For $0<w<\pi$,
\begin{equation}
 g'(w)=2\sin\frac{w}{2}\left(2\sin\frac{w}{2}-w\cos\frac{w}{2}\right)>0,
\end{equation}
because $\tan(w/2)>w/2$, and $g(0)=0$.  For $w\geq\pi$, if $\sin w\leq0$ then $g(w)>0$ immediately; if $\sin w>0$, then $2+\cos w\geq1>3/\pi\geq3\sin w/w$, again giving $g(w)>0$.  Hence $h(w)>0$ for every $w\neq0$.  Averaging over any projected-angle distribution satisfying $\Pr(\theta\neq0)>0$ therefore gives $\left.\partial G_z/\partial z\right|_X>0$ for $X>0$.

\begin{acknowledgments}
This work was supported by the National Natural Science Foundation of China (62375215), the Shaanxi Province Innovation Capability Support Plan (2024RS-CXTD-07), the Key R\&D Program of Shaanxi Province (2025PT-ZCK-04), and the 111 Project (B14040).
\end{acknowledgments}

\section*{Data availability}
The raw and processed data, together with the analysis and plotting code supporting the findings of this study, will be made publicly available upon publication.

\bibliography{pra_regular_article_refs}

@article{Neil1997,
  author = {Neil, M. A. A. and Ju\v{s}kaitis, R. and Wilson, T.},
  title = {Method of obtaining optical sectioning by using structured light in a conventional microscope},
  journal = {Opt. Lett.},
  volume = {22},
  pages = {1905--1907},
  year = {1997},
  doi = {10.1364/OL.22.001905}
}

@article{Cuccia2009,
  author = {Cuccia, D. J. and Bevilacqua, F. and Durkin, A. J. and Ayers, F. R. and Tromberg, B. J.},
  title = {Quantitation and mapping of tissue optical properties using modulated imaging},
  journal = {J. Biomed. Opt.},
  volume = {14},
  pages = {024012},
  year = {2009},
  doi = {10.1117/1.3088140}
}

@article{Wells1969,
  author = {Wells, W. H.},
  title = {Loss of resolution in water as a result of multiple small-angle scattering},
  journal = {J. Opt. Soc. Am.},
  volume = {59},
  pages = {686--691},
  year = {1969},
  doi = {10.1364/JOSA.59.000686}
}

@article{Yura1971,
  author = {Yura, H. T.},
  title = {Small-Angle Scattering of Light by Ocean Water},
  journal = {Appl. Opt.},
  volume = {10},
  number = {1},
  pages = {114--118},
  year = {1971},
  doi = {10.1364/AO.10.000114}
}

@article{Kuga1985,
  author = {Kuga, Y. and Ishimaru, A.},
  title = {Modulation transfer function and image transmission through randomly distributed spherical particles},
  journal = {J. Opt. Soc. Am. A},
  volume = {2},
  pages = {2330--2336},
  year = {1985},
  doi = {10.1364/JOSAA.2.002330}
}

@article{Kuga1986,
  author = {Kuga, Y. and Ishimaru, A.},
  title = {Modulation transfer function of layered inhomogeneous random media using the small-angle approximation},
  journal = {Appl. Opt.},
  volume = {25},
  pages = {4382--4385},
  year = {1986},
  doi = {10.1364/AO.25.004382}
}

@article{Benke1990,
  author = {Benke, K. K. and McKellar, B. H. J.},
  title = {Modulation transfer function of photographic emulsion: the small-angle approximation in radiative transfer theory},
  journal = {Appl. Opt.},
  volume = {29},
  pages = {151--156},
  year = {1990},
  doi = {10.1364/AO.29.000151}
}

@article{McLean1991,
  author = {McLean, J. W. and Voss, K. J.},
  title = {Point spread function in ocean water: comparison between theory and experiment},
  journal = {Appl. Opt.},
  volume = {30},
  pages = {2027--2030},
  year = {1991},
  doi = {10.1364/AO.30.002027}
}

@article{Swanson2001,
  author = {Swanson, N. L. and Gehman, V. M. and Billard, B. D. and Gennaro, T. L.},
  title = {Limits of the small-angle approximation to the radiative transport equation},
  journal = {J. Opt. Soc. Am. A},
  volume = {18},
  pages = {385--391},
  year = {2001},
  doi = {10.1364/JOSAA.18.000385}
}

@article{Kim2004,
  author = {Kim, A. D.},
  title = {Transport theory for light propagation in biological tissue},
  journal = {J. Opt. Soc. Am. A},
  volume = {21},
  pages = {820--827},
  year = {2004},
  doi = {10.1364/JOSAA.21.000820}
}

@article{Dolin2013,
  author = {Dolin, L. S.},
  title = {Theory of lidar method for measurement of the modulation transfer function of water layers},
  journal = {Appl. Opt.},
  volume = {52},
  pages = {199--207},
  year = {2013},
  doi = {10.1364/AO.52.000199}
}

@article{Gray2012,
  author = {Gray, Deric J.},
  title = {Order-of-scattering point spread and modulation transfer functions for natural waters},
  journal = {Appl. Opt.},
  volume = {51},
  pages = {6753--6764},
  year = {2012},
  doi = {10.1364/AO.51.006753}
}

@book{Ishimaru1978,
  author = {Ishimaru, A.},
  title = {Wave Propagation and Scattering in Random Media},
  publisher = {Academic Press},
  address = {New York},
  year = {1978}
}

@article{Cuccia2005,
  author = {Cuccia, D. J. and Bevilacqua, F. and Durkin, A. J. and Tromberg, B. J.},
  title = {Modulated imaging: quantitative analysis and tomography of turbid media in the spatial-frequency domain},
  journal = {Opt. Lett.},
  volume = {30},
  pages = {1354--1356},
  year = {2005},
  doi = {10.1364/OL.30.001354}
}

@article{Kanick2014,
  author = {Kanick, S. C. and McClatchy, D. M. and Krishnaswamy, V. and Elliott, J. T. and Paulsen, K. D. and Pogue, B. W.},
  title = {Sub-diffusive scattering parameter maps recovered using wide-field high-frequency structured light imaging},
  journal = {Biomed. Opt. Express},
  volume = {5},
  pages = {3376--3390},
  year = {2014},
  doi = {10.1364/BOE.5.003376}
}

@article{Hu2017,
  author = {Hu, D. and Lu, R. and Ying, Y.},
  title = {Finite element simulation of light transfer in turbid media under structured illumination},
  journal = {Appl. Opt.},
  volume = {56},
  pages = {6035--6042},
  year = {2017},
  doi = {10.1364/AO.56.006035}
}

@article{Gioux2019,
  author = {Gioux, S. and Mazhar, A. and Cuccia, D. J.},
  title = {Spatial frequency domain imaging in 2019: principles, applications, and perspectives},
  journal = {J. Biomed. Opt.},
  volume = {24},
  pages = {071613},
  year = {2019},
  doi = {10.1117/1.JBO.24.7.071613}
}

@article{Shapiro2008,
  author = {Shapiro, J. H.},
  title = {Computational ghost imaging},
  journal = {Phys. Rev. A},
  volume = {78},
  pages = {061802},
  year = {2008},
  doi = {10.1103/PhysRevA.78.061802}
}

@article{Zhao2022,
  author = {Zhao, X. and Nie, X. and Yi, Z. and Peng, T. and Scully, M. O.},
  title = {Imaging through scattering media via spatial--temporal encoded pattern illumination},
  journal = {Photonics Res.},
  volume = {10},
  pages = {1689--1694},
  year = {2022},
  doi = {10.1364/PRJ.456156}
}

@article{Wang2025Implicit,
  author = {Wang, C. and He, Y. and Wang, Y. and Chen, H. and Zheng, H. and Liu, J. and Zhou, Y. and Xu, Z.},
  title = {Implicit representation optimization method for enhanced ghost imaging through scattering medium},
  journal = {Opt. Lasers Eng.},
  volume = {195},
  pages = {109262},
  year = {2025},
  doi = {10.1016/j.optlaseng.2025.109262}
}

@article{Wang2026Graph,
  author = {Wang, C. and Lu, H. and Han, Y. and Zheng, H. and He, Y.},
  title = {Graph-Regularized Estimation for Inverse Problems in Computational Optical Imaging},
  journal = {IEEE Trans. Image Process.},
  volume = {35},
  pages = {9314--9326},
  year = {2026},
  doi = {10.1109/TIP.2026.3726365}
}

@article{Zardecki1984,
  author = {Zardecki, A. and Gerstl, S. A. W. and Embury, J. F.},
  title = {Multiple scattering effects in spatial frequency filtering},
  journal = {Appl. Opt.},
  volume = {23},
  pages = {4124--4131},
  year = {1984},
  doi = {10.1364/AO.23.004124}
}

@article{Bruscaglioni1993,
  author = {Bruscaglioni, P. and Donelli, P. and Ismaelli, A. and Zaccanti, G.},
  title = {Monte Carlo calculations of the modulation transfer function of an optical system operating in a turbid medium},
  journal = {Appl. Opt.},
  volume = {32},
  pages = {2813--2824},
  year = {1993},
  doi = {10.1364/AO.32.002813}
}

@article{Jaffe1995,
  author = {Jaffe, J. S.},
  title = {Monte Carlo modeling of underwater-image formation: validity of the linear and small-angle approximations},
  journal = {Appl. Opt.},
  volume = {34},
  pages = {5413--5421},
  year = {1995},
  doi = {10.1364/AO.34.005413}
}

@article{Sassaroli1998,
  author = {Sassaroli, A. and Blumetti, C. and Martelli, F. and Alianelli, L. and Contini, D. and Ismaelli, A. and Zaccanti, G.},
  title = {Monte Carlo procedure for investigating light propagation and imaging of highly scattering media},
  journal = {Appl. Opt.},
  volume = {37},
  pages = {7392--7400},
  year = {1998},
  doi = {10.1364/AO.37.007392}
}

@article{Rogers2019,
  author = {Rogers, G.},
  title = {Transmission point spread function of a turbid slab},
  journal = {J. Opt. Soc. Am. A},
  volume = {36},
  pages = {1617--1623},
  year = {2019},
  doi = {10.1364/JOSAA.36.001617}
}

@article{Gan1999,
  author = {Gan, X. and Gu, M.},
  title = {Effective point-spread function for fast image modeling and processing in microscopic imaging through turbid media},
  journal = {Opt. Lett.},
  volume = {24},
  pages = {741--743},
  year = {1999},
  doi = {10.1364/OL.24.000741}
}

@article{LeBihan2009,
  author = {{Le Bihan}, N. and Margerin, L.},
  title = {Nonparametric estimation of the heterogeneity of a random medium using compound Poisson process modeling of wave multiple scattering},
  journal = {Phys. Rev. E},
  volume = {80},
  pages = {016601},
  year = {2009},
  doi = {10.1103/PhysRevE.80.016601}
}

@book{Goodman2017,
  author = {Goodman, J. W.},
  title = {Introduction to Fourier Optics},
  edition = {4},
  publisher = {W. H. Freeman},
  address = {New York},
  year = {2017}
}

@book{Kingman1992,
  author = {Kingman, J. F. C.},
  title = {Poisson Processes},
  series = {Oxford Studies in Probability},
  volume = {3},
  publisher = {Oxford University Press},
  address = {Oxford},
  year = {1992},
  doi = {10.1093/oso/9780198536932.001.0001}
}

\end{document}